\documentclass[aps,prl,twocolumn,superscriptaddress]{revtex4-2}
\usepackage{graphicx}   
\usepackage{bm}        
\usepackage{comment}
\usepackage{amssymb}   
\usepackage{amsmath}
\usepackage{tabularx}
\usepackage{tabulary}
\usepackage[usenames,dvipsnames]{color}
\usepackage[normalem]{ulem}

\begin{document}

\preprint{APS/123-QED}

  \title{
   Multi-messenger detection of black hole binaries in dark matter spikes}
   \author{Fani Dosopoulou}
   \email{dosopoulouf@cardiff.ac.uk}
    \affiliation{School of Physics and Astronomy, Cardiff University, Cardiff, CF24 3AA, United Kingdom}
\author{Joseph Silk}
\affiliation{William H. Miller III Department of Physics and Astronomy, The Johns Hopkins University, Baltimore, Maryland 21218, USA}
\affiliation{Institut d’Astrophysique de Paris, UMR 7095 CNRS and UPMC, Sorbonne Universite´, F-75014 Paris, France}

\begin{abstract}
 We investigate the inspiral of a high mass-ratio black hole binary located in the nucleus of a galaxy, where the primary central black hole is surrounded by a dense dark matter spike formed through accretion during the black hole growth phase. Within this spike, dark matter undergoes strong self-annihilation, producing a compact source of $\gamma$-ray radiation that is highly sensitive to spike density, while the binary emits gravitational waves at frequencies detectable by LISA. As the inspiralling binary interacts with the surrounding dark matter particles, it alters the density of the spike, thereby influencing the $\gamma$-ray flux from dark matter annihilation. We demonstrate that the spike self-annihilation luminosity decreases  by $10\%$ to $90\%$ of its initial value, depending on the initial density profile and binary mass ratio, as the binary sweeps through the LISA band. This presents a new opportunity to indirectly probe dark matter  through multi-messenger observations of galactic nuclei.
\end{abstract}
\maketitle

\noindent
{\it Introduction.}~The primary targets for the Laser Interferometer Space Antenna (LISA) are gravitational waves (GWs) generated by the inspiral and coalescence of supermassive black hole (SMBH) binaries \cite{AmaroSeoane:2012km,2017arXiv170200786A,2023LRR....26....2A,Barausse:2014tra}. Among the most promising sources are 
extreme-mass ratio inspirals (EMRIs)
in which a stellar-mass compact object gradually inspirals into a SMBH \cite{2023LRR....26....2A,2024arXiv240207571C}. These systems are expected to provide unprecedented insights into strong-field gravity, black hole demographics, and the dynamics of galactic nuclei \cite{2009CQGra..26i4034G,2010PhRvD..81j4014G,2011PhRvD..83d4036S,2023arXiv231213028K,2024arXiv241103436D}. Moreover, their signals will serve as powerful probes of the nuclear environments in which they reside, offering a unique window into the interplay between compact objects and their surroundings \cite{2009CQGra..26i4034G,2015JPhCS.610a2002A,2024arXiv241103436D,2025A&A...693A..22M}.

When a SMBH grows adiabatically within a cuspy dark matter (DM) halo, it induces the formation of a high-density central cusp of DM, known as a spike \cite{Gondolo:1999ef,Sadeghian:2013laa}. Such DM spikes are particularly interesting for indirect detection efforts, as the self-annihilation of DM particles (e.g., weakly interacting massive particles, or WIMPs) in these regions is expected to produce $\gamma$-rays, potentially detectable by instruments like the Fermi Large Area Telescope (Fermi-LAT) \cite{2005PhRvD..72j3502B,2012AnP...524..479B}.  The gravitational interaction between the binary and the DM spike can modify the emitted GW signal, providing a novel avenue to constrain the properties of DM and test its existence in dense astrophysical environments \cite[e.g.,][]{Eda:2013gg,Eda:2014kra,2025arXiv250113601D,Yue:2017iwc,Macedo:2013qea,Yue:2019ndw,Cardoso:2019rou,2020PhRvD.102h3006K,2022PhRvD.105f3029B,2023PhRvD.107h3003B,2024arXiv240519240Z,2024PhRvD.110h3027D}.

The gravitational influence of a binary within a DM spike can also perturb the spike density profile,  affecting the $\gamma$-ray flux from DM annihilation. This change in the $\gamma$-ray emission could serve as an indirect signature of the binary black hole (BBH)  impact on the DM distribution. By correlating observed GW signals from BBH mergers with changes in $\gamma$-ray emission, it may be possible to probe the properties of DM and test the self-annihilation hypothesis in regions of high DM density \cite{2024JCAP...09..005A,2024NuPhB100316487B}. 

In this work, we investigate the correlation between GWs emitted by BBH systems and the reduction in $\gamma$-ray emission from self-annihilating DM in DM spikes. We model the dynamics of BBHs embedded in DM spikes, accounting for the gravitational perturbation of the spike by the binary system. Using this framework, we calculate the expected change in $\gamma$-ray flux due to the disruption of the DM density profile. This approach  provides a novel method to simultaneously probe the properties of DM and the dynamics of BBH systems. It also establishes a new multi-messenger framework for studying dense astrophysical environments.

\bigskip
\noindent
{\it Methods.}~We use $N$-body simulations to model the response of a DM spike to the presence of a massive BBH that is evolving due to energy loss by GW emission.
We first define several parameters for the masses of the binary components
and the DM distribution, and note some approximations that we make. 
We  denote the mass of the
central black hole by $M$ and the mass of the smaller black hole by $m$, and the binary mass ratio $q=m/M$.  The binary is embedded within a DM spike, formed in a DM halo as a consequence of  the adiabatic growth of a central black hole \cite{Gondolo:1999ef,Sadeghian:2013laa}.

A ‘spiked’ DM density cusp is predicted to form inside
$r_{\rm sp} \approx 0.2r_{\rm h}$ with a power law form of slope $\gamma_{\rm sp}$, where $r_{\rm h}$ is the radius of influence of the SMBH, defined as the radius containing a mass $2M_{ }$ and $\gamma_{\rm sp} \approx 1.5-2.5$ \cite{2020PhRvD.102h3006K,2017ApJ...840...43A,2024arXiv240601705C}.
We consider a DM spike with a power law initial density profile  $\rho_{\rm sp}(r)\propto {r} ^{-\gamma}$.
 Below we assume that the binary dynamics is not influenced by DM and that the DM cusp behaves as a fully collisionless system. This makes our results  independent of the chosen density scaling or normalisation, i.e., they only depend on binary parameters and not on the absolute value of density.
 These assumptions are well justified since, in the relevant regime, the evolution of the binary is dominated by GW energy loss and DM particle-particle interactions are unimportant.

We generate the initial particle positions under the assumption that the DM is spherically symmetric and follows the power law density profile given above. This profile is truncated at a radius \( r = r_{\rm cut} \) by multiplying the spike density \( \rho_{\rm sp} \) by the function:
$
{2}/[{\cosh(r/r_{\rm cut}) + \text{sech}(r/r_{\rm cut})}].
$
This truncation ensures a smooth transition in the density profile at the cutoff radius \( r_{\rm cut} \).
 The  particle velocities are then generated  self-consistently in the potential of the central black from a numerically computed distribution function.


After the initial particle velocities and positions are assigned, we place a second BH on a circular orbit around the primary BH at a distance such that the GW merger time is $\sim~10\rm \:yr$.
We then integrate the particle trajectories  in the potential of the two BHs. 
We assume that any particle that approaches 
the primary (secondary) BH within a distance $4 GM/c^2$ ($4 Gm/c^2$) is captured and we remove it  from the system. 
 This choice of capture radius serves as a relativistically motivated proxy 
for the region where strong-field effects dominate and particle orbits become unstable.
The integration is done using the code {\tt RAGA} \citep{2015MNRAS.446.3150V}. Particle trajectories are computed independently and in parallel using  the  eighth-order Runge–Kutta method {\tt dop853} \cite{1988ZaMM...68..260S}.
The effect of particle-particle interactions is not accounted for in the calculation as the DM is considered to be a fully collisionless system.

\begin{figure}        
\includegraphics[height=0.66\columnwidth]{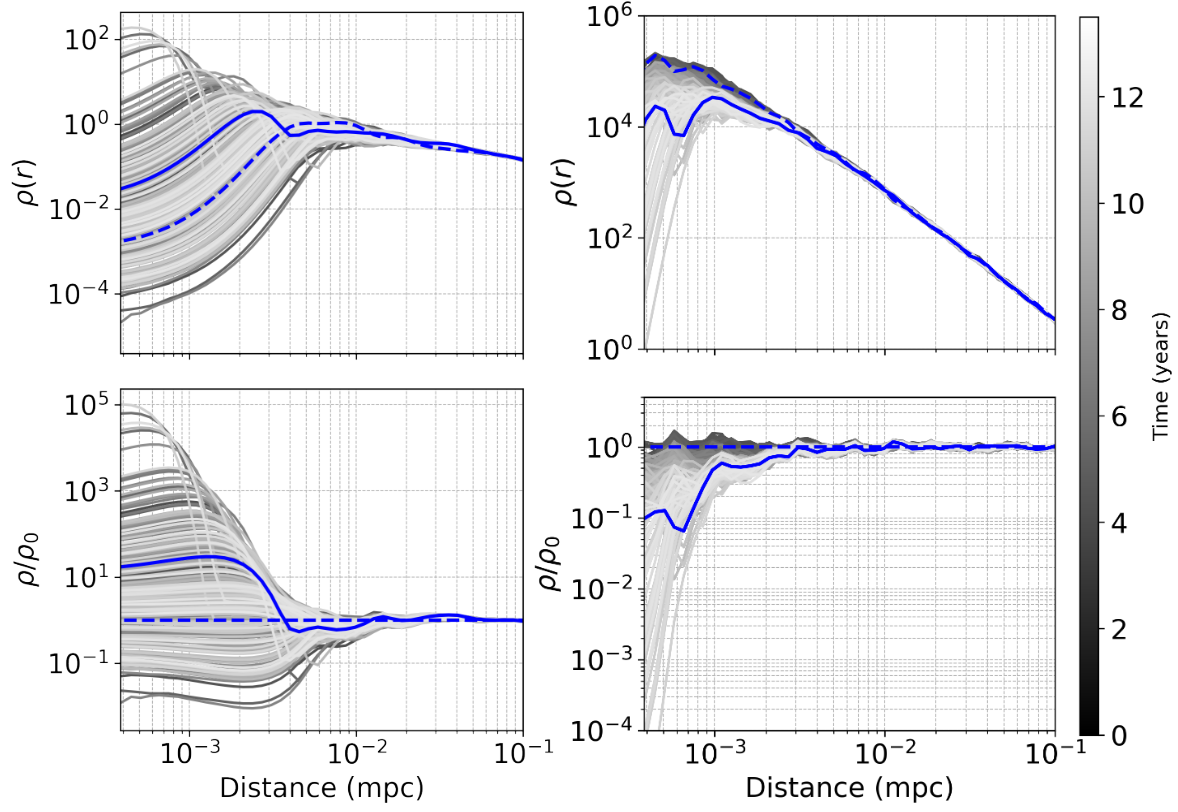}      
    \caption{The time evolution of the DM spike density $\rho(r)$ due to the BBH inspiral is shown for $\gamma=0.5$ (left panels) ane $\gamma=7/3$ (right panels). The x-axis lower bound is the radius
    $4GM/c^2$ below which particles are captured by the primary BH.    
    The blue dashed lines represent the initial density profile of the DM spike, while the solid blue lines correspond to the final density just after merger. The upper panel illustrates the density change as a function of time, indicated by the grey color map on the right. The lower panel presents the density change normalized by the initial density, $\rho(t=0)$.}
    \label{fig:allplots}
\end{figure}

We assume that the binary evolution is dominated by GW energy loss, and 
compute the evolution of the binary semi-major axis using the orbit-averaged evolution equations from \citet{1964PhRv..136.1224P}. 
The integration is halted when the binary reaches coalescence.
 Any  dynamical effect
of  DM on the evolution of the binary is neglected and we discuss why this is a reasonable choice  in the Supplementary Material (SM).  These effects include DM accretion onto the BHs and dynamical friction. Although these are  secondary in their influence on both the evolution of the binary and of the DM density, they can leave a detectable dephasing of the GW signal which  has been previously demonstrated  \cite[e.g.,][]{2020PhRvD.102h3006K,2022PhRvD.105d3009C,2024PhRvD.110h3027D}. As noted above, a direct consequence   of  a negligible impact of DM interactions on the binary and of the collisionless nature of DM is that the
 density response is inherently scale-free because the evolution of DM particles depends only on the changing binary potential.

\bigskip
\noindent
{\it Annihilation flux.}
The annihilation flux (i.e., photons per unit energy per unit area per unit time) from an astrophysical region  is represented as a product of two quantities. The first
depends on particle physics and the second, called
“astrophysical factor” $J$, is related to the DM spatial density $\rho(r)$:
$\Phi = {\frac{1}{2}}\frac{\langle \sigma v \rangle}{m_{\chi}^2} {\frac{1}{D^2}}\sum_f \frac{dN_{\gamma}^f}{dE} \times \bar{J}$,
where \( \langle \sigma v \rangle \) is the thermally averaged annihilation cross-section, \( m_{\chi} \) is the mass of the DM particle, \( \frac{dN_{\gamma}^f}{dE} \) is the $\gamma$-ray spectrum per annihilation into a final state \( f \), and $D$ is the distance to the source.

We can write the average $\bar{J}$ factor in terms of the “vicinity” of the black hole and the “background” from outside this vicinity \citep{2007PhRvD..76j3532V}:
\begin{equation}\label{Jflux}
\bar{J}=\int_{0}^{R_{\rm max}} \rho^2(r)r^{2}dr+\bar{J}_{\rm bkg}
\end{equation}
where $R_{\rm max}$ is the radius where the spike density drops to zero.  In what follows we assume that the first term due to the spike is dominant and neglect the contribution of $\bar{J}_{\rm bkg}$.
Eq.~\ref{Jflux} shows that
the total annihilation flux scales as $\rho^2$, and it is therefore very sensitive to the DM density. Also,  it is evident that if $\gamma < 1.5$, then most part of the flux comes from the outer boundary of
the corresponding region $0<r<R_{\rm max}$, and if $\gamma > 1.5$, the integral is
dominated by the flux produced near the inner boundary. This implies that if the power-law index everywhere outside $r_{\rm h}$ is
greater than $1.5$, then the annihilation area outside $r_{\rm h}$, as well as the background contributes little to the total annihilation flux.
Consequently, in such cases, the total flux becomes highly sensitive to any variations in the DM density near the SMBH. We should therefore expect  our results to depend  on both the initial binary parameters and the density profile slope of the cusp. In what follows, we investigate the impact of the BBH inspiral on the $\gamma$-ray flux generated by self-annihilation processes in the DM spike, and examine its dependence on $\gamma$ and $q$.

\bigskip

\noindent
{\it Results.}
\label{results}
\begin{figure}
\includegraphics[width=0.45\textwidth]{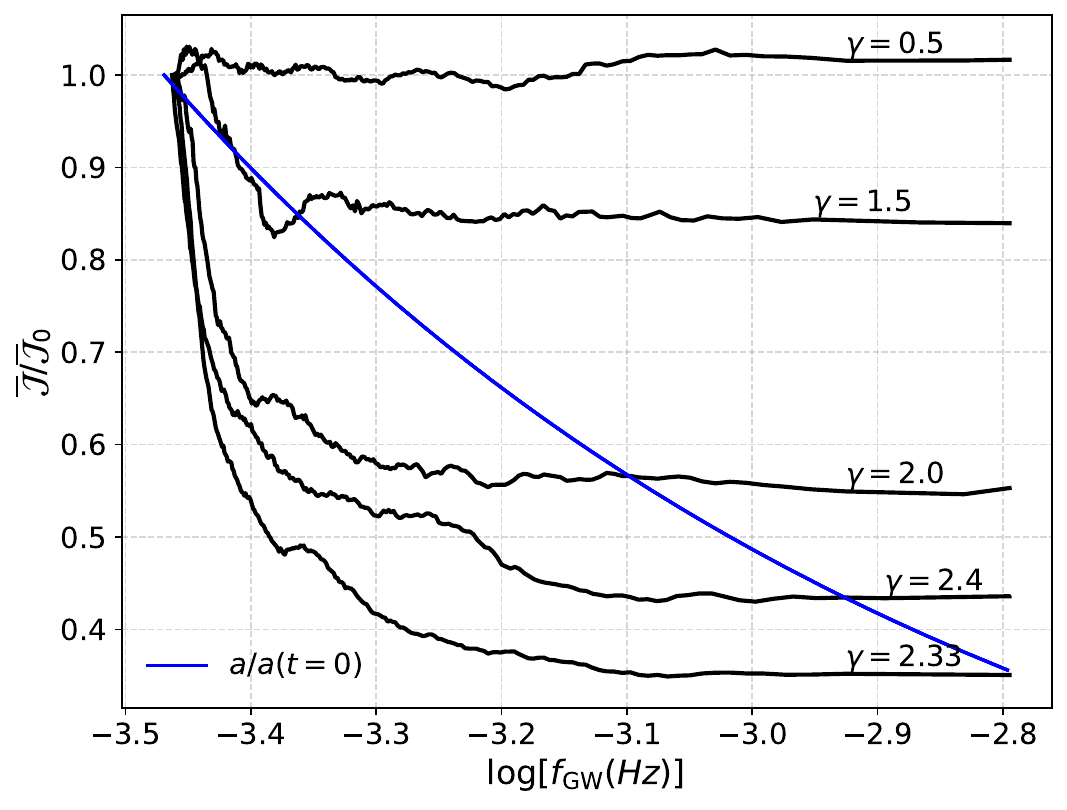}
\includegraphics[width=0.45\textwidth]{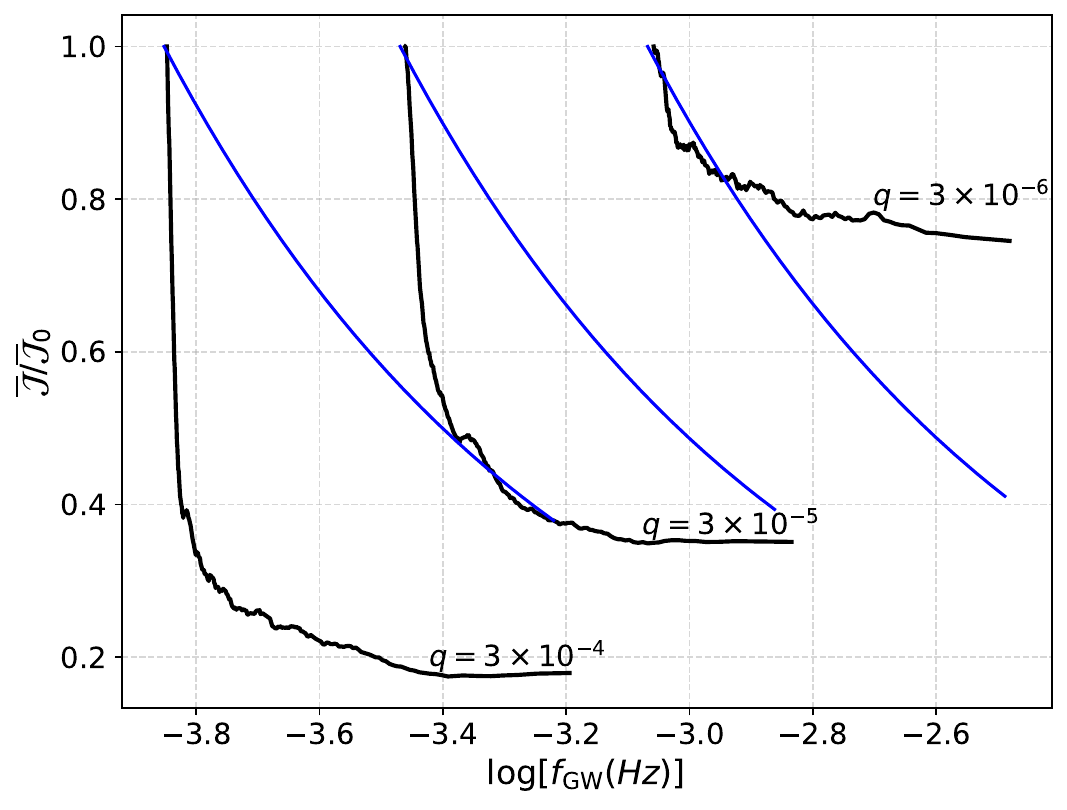}
    \caption{The $\gamma$-ray flux is shown as a function of the GW frequency, $f_{\rm GW}$, of the BBH for different values of $\gamma$ (top) and different mass ratios $q$ with $\gamma=7/3$ (bottom). In both plots, the flux is normalized by its initial value, $\mathcal{J}_{0}\equiv \mathcal{J}(t=0)$. The blue lines represent the time evolution of the binary semi-major axis, normalized by its initial value, $a/a_0$.}
    \label{fig:combined}
\end{figure}
We consider a binary system with a primary black hole mass of $M = 4 \times 10^6 \, M_\odot$, a mass ratio of $q = 3 \times 10^{-5}$, initial semi-major axis $ a_0\equiv a(t=0)= 2.5 \times 10^{-3} \, \text{mpc}$, and initial eccentricity $e_0 = 0$. The binary initial GW frequency is $8.9\times 10^{-3}$Hz; i.e., it is within the LISA frequency band $0.1\rm mHz\ - 1 \rm Hz$ \cite{AmaroSeoane:2012km,2017arXiv170200786A,2023LRR....26....2A,Barausse:2014tra}. For a circular binary, the GW frequency is related to the orbital angular frequency, $\Omega_{\rm orb}=\sqrt{G[M+m]/a}$, through the simple relation $f_{\rm GW}=\Omega/\pi$ \cite{2003ApJ...598..419W},
while the  strain amplitude of the GW signal evolves as $h\propto f_{\rm GW}^{2/3}$.
Using these initial conditions, we investigate the dependence of the produced $\gamma$-ray flux on various DM spike density profiles with slopes $\gamma = (0.5, 1.5, 2.0, 7/3, 2.4)$ and we choose a truncation radius $r_{\rm cut}=0.1\rm \: mpc$. We note that a $\gamma = 7/3$ DM spike is expected to form in the center of a halo with an initial profile $\rho \propto r^{-1}$, such as for the NFW profile \cite{Gondolo:1999ef,2020PhRvD.102h3006K,2017ApJ...840...43A,2024arXiv240601705C}.  For $\gamma = 7/3$, we also explore how the results are affected by the binary mass ratio, running two additional simulations, with $M=4 \times 10^6 \: M_\odot$, $q = 3 \times 10^{-6}$ and  $a_0=1.4\times 10^{-3}$mpc, and $q = 3 \times 10^{-4}$ and $a_0=4.5\times 10^{-3}$mpc. The initial orbits were set such that the GW insipiral timescale was $\simeq 12$yr.

During the BBH inspiral, some fraction of the energy loss is transferred to surrounding DM particles, ejecting or displacing them from the cusp. The code {\tt RAGA} models the effect of the central SMBH binary on the DM cusp distribution, the scattering of DM particles and the evolution of the binary orbit due to GWs. In Fig.~\ref{fig:allplots}, we present the evolution of the DM spike density profile as a function of the time until merger, along with the density change. The scattering of DM particles by the black holes influences the DM spike density at radii $r < a_0$.  For $\gamma=7/3$, the BBH inspiral leads to a progressive decrease in the DM density over time, whereas for $\gamma=0.5$, we observe larger variations of the central density within $a_0$.

In Fig.~\ref{fig:combined}, we apply Eq.~\ref{Jflux} to examine how the temporal evolution of the DM spike density distribution affects the total annihilation flux. We present the $\gamma$-ray flux as a function of the GW frequency of the BBH system. In addition, we show the results for the different values of the mass ratio.  As a reference for the  time evolution of the binary orbit, we also show the variation of the binary semi-major axis. As discussed above, we expect that for $\gamma \lesssim 1.5$, the time variation of the DM spike density distribution has a small effect on the produced $\gamma$-ray flux. Accordingly, 
although the $\gamma=0.5$ case shows the larger  central density variation with time (see Fig.~\ref{fig:allplots}), we see a negligible time dependence of the flux. 
For $\gamma=1.5,$ the flux decreases to 
$0.8\mathcal{J}_{0}$
by the end of the simulation, with $\mathcal{J}_{0}\equiv \mathcal{J}(t=0)$.
On the other hand,  for $\gamma > 1.5$, the total annihilation flux is highly sensitive to changes in the DM density in the innermost regions, resulting in a reduction of the total flux to   $0.3\mathcal{J}_{0}$ for $\gamma=7/3$.

 In Fig.~\ref{fig:combined}, we also explore the dependence of the $\gamma$-ray flux change with respect to the binary mass ratio. For the same density profile of the DM spike, the  annihilation flux at the end of the simulation decreases with the mass-ratio of the black hole binary, ranging from $\simeq 0.1\mathcal{J}_{0}$ for $q=3\times 10^{-4}$ to $\simeq 0.9\mathcal{J}_{0}$ for $q=3\times 10^{-6}$. This dependence on $q$ is expected, as the disruption of the DM spike density distribution 
is caused by 
 scattering of the DM particles off the secondary black hole and 
it is therefore more pronounced for a heavier secondary.

\begin{figure}
\includegraphics[width=0.5\textwidth]{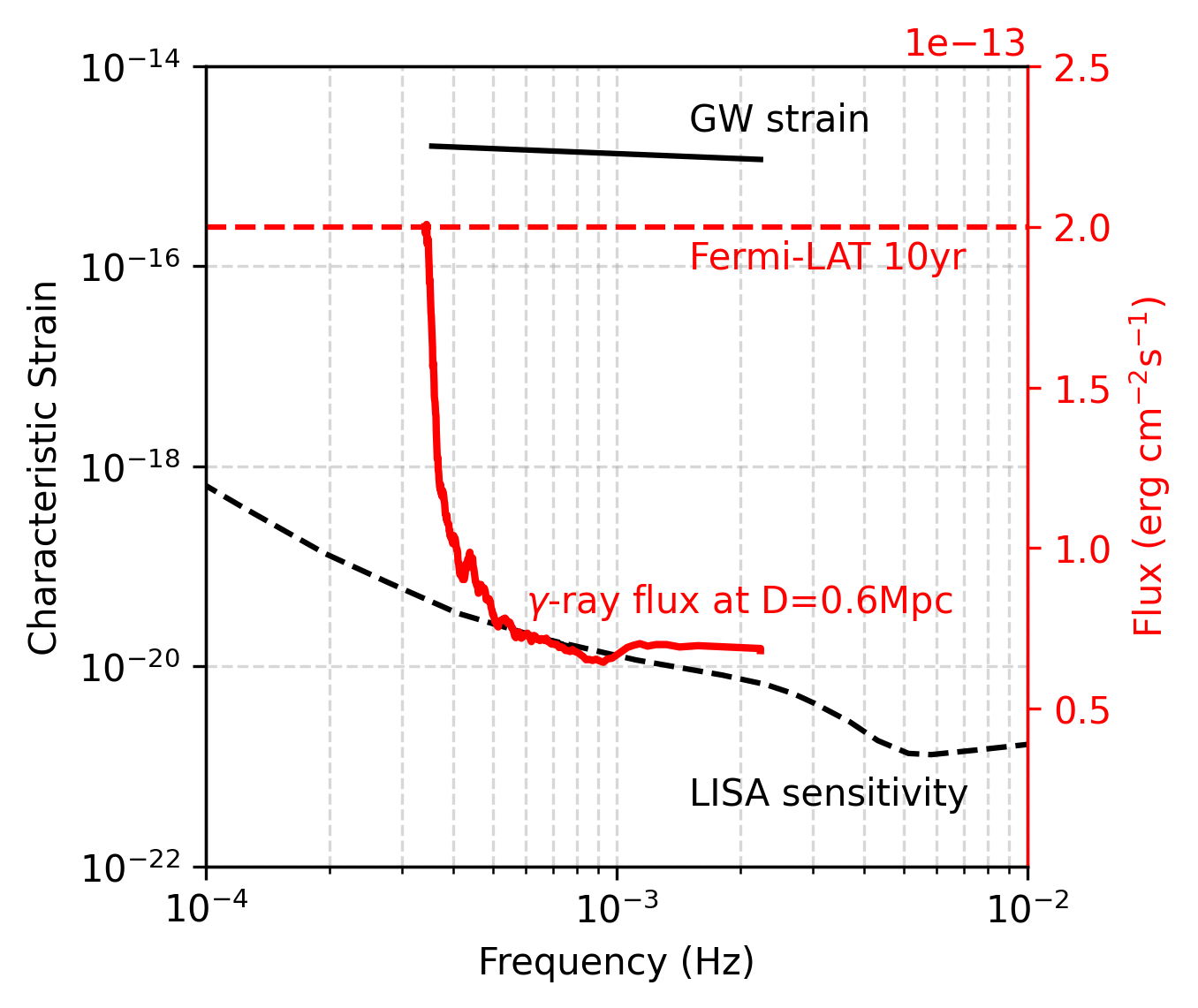}
    \caption{
   The red solid curve show  a representative value of the $\gamma$-ray flux at 2 Gev for  $\gamma=7/3$, $q=3\times 10^{-5}$ and   $M=4\times 10^6 \:M_\odot$. 
    The red-dashed line is the Fermi-LAT  flux sensitivity at 2 Gev, i.e., the minimum flux needed to get a n-standard-deviation detection from a point-like $\gamma$-ray source at this energy, estimated for a data taking of duration $10\: \rm yr$.
   The initial source luminosity has been set to a value typical for a canonical DM spike in a Milky-Way like galaxy, and its distance
 ($D=0.65$Mpc) is such that the flux is initially at the Fermi sensitivity limit. The black lines show the characteristic strain of the BBH  (solid) and the LISA sensitivity curve from \cite{2019CQGra..36j5011R} (dashed).
   }
    \label{fermi}
\end{figure}

\bigskip
\noindent
{\it Detectability.}~Our analysis suggests that the correlation between GWs from BBH inspirals and the suppression of $\gamma$-ray emission from DM annihilation presents a novel multi-messenger approach to probing DM. However, whether these flux variations are detectable by current or future $\gamma$-ray observatories depends on several factors. These include the initial $\gamma$-ray flux level, the magnitude of the flux suppression, the time-scale of the variation, and the sensitivity of the telescope. 
Detecting $\gamma$-ray flux variations at large distances is particularly important for enhancing the probability of observing a binary in the LISA band.

We consider the detectability of the signal. The predicted differential flux from self-annihilating DM is subject to several uncertainties, particularly in the particle mass and interaction cross-sections. For DM annihilating into \( b\bar{b} \), the typical flux for a canonical adiabatic spike at the Galactic Center is estimated to be \( \sim 10^{-9} \,{\rm erg/s/cm^2} \) at 2 GeV, where the energy spectrum reaches its peak \cite{2014PhRvL.113o1302F,2016PDU....12....1D,2024arXiv240601705C}. The Fermi LAT detection limit for a $\gamma$-ray source at this energy is approximately \( 10^{-13} \,{\rm erg/s/cm^2} \) \cite{2023arXiv230519690B}. Given a galactocentric distance of \( 8 \) kpc, this suggests that a source of such intensity could still be detectable by Fermi LAT at distances up to \( \sim 1 \) Mpc, extending  into the Local Group. The representative $\gamma$-ray signal of a BBH that initially is just above the Fermi LAT sensitivity is shown in Fig.~\ref{fermi}. We see  that with current instruments,
a $\gamma$-ray detection is unlikely to extend beyond the Local Group.

In parallel, LISA might be expected to detect several tens to a few hundred EMRIs over its operational lifetime, primarily at low redshifts (\( z \lesssim 1 \)) \cite{2009CQGra..26i4034G}. 
Fig.~\ref{fermi} shows that for a detectable $\gamma$-ray source,
the GW of the binary can be clearly identified in LISA.
However, the detection rate is low as the overall merger rate distribution is projected to peak at  \( z \sim 0.2 \) \cite{2009CQGra..26i4034G}. But, due to existing uncertainties in event rates, it remains unclear how many of these inspirals will occur in sufficiently nearby environments where the associated $\gamma$-ray signal from a pronounced DM spike could be resolved. Moroever,  LISA  is expected to reach sky localization of EMRIs to within a few square
degrees. This will allow a $\gamma$-ray telescope to focus its observations on a well-defined region instead of scanning the whole sky.
This reduces the background noise and increases sensitivity to the faint $\gamma$-ray flux.

Future $\gamma$-ray telescopes with enhanced sensitivity will be essential for extending the detectable range, thereby increasing the likelihood of observing a multimessenger signal. Although challenging, the potential scientific impact of such a detection would be profound, as it would constitute a decisive indirect confirmation of the existence of DM.

 We have implicitly assumed a generic self-annihilating dark matter candidate, such as a WIMP.  For non-annihilating candidates, such as axions or asymmetric DM, the associated $\gamma$-ray signal would be suppressed or absent, and the multi-messenger signatures discussed here would not apply. Our approach remains relevant for any scenario in which dark matter self-annihilation is efficient in high-density environments.

\bigskip
\noindent
{\it Conclustions.}~We have investigated the response of a DM spike to the presence of a massive BBH system that evolves due to energy loss through GW emission. We have shown that the scattering of DM particles during the BBH inspiral leads to changes in the DM spike density in the inner regions of the spike. We have studied how the temporal evolution of the DM spike density distribution affects the time variation of the total annihilation flux and examined its dependence on both the density profile slope of the DM spike and the binary mass ratio.

 In DM spikes with a density slope shallower than $\gamma<1.5,$ changes in the inner region of the spike have little effect on the total annihilation flux. 
 For steeper  DM spikes  the annihilation flux is dominated by the region near the inner boundary, making the total flux  sensitive to any reduction in DM density due to an inspiraling BBH.

 As the BBH inspiral progresses, this leads to a significant decrease in the $\gamma$-ray flux, demonstrating a correlation between the GWs emitted and the $\gamma$-ray emission from self-annihilating DM.
The annihilation flux can experience a reduction of up to  one-order of magnitude, indicating that dense inner regions of the DM spike are particularly sensitive to BBH-induced scattering effects.

While our study focuses on the evolution of DM spikes under the influence of a BBH system, additional astrophysical processes, particularly those involving baryons, could significantly alter the spike structure. Gravitational interactions with stars may heat the DM, reducing its central density over time, with some models suggesting a long-term evolution toward a shallower slope of \( \gamma \sim 3/2 \) \cite{2004PhRvL..93f1302G}.  Other factors, such as past mergers \cite{2002PhRvL..88s1301M,2014PhRvL.113o1302F,2018PhRvD..98b3536K}, supernova-driven gravitational potential fluctuations \cite{2022MNRAS.513.3458B}, DM self-interactions \cite{2014PhRvL.113o1302F,2024PhRvL.133b1401A,2014PhRvD..89b3506S}, or unseen compact objects \cite{2014PhRvL.113o1302F}, could also reshape the spike. Additionally, the formation history of the SMBH plays a crucial role, with rapid formation leading to shallower spikes and gradual accretion allowing steeper structures to develop. 
 Finally,  compact dark matter structures such as mounds \cite{2024arXiv240408731B} and crests \cite{2007PhRvD..75d3517M} could likewise be affected. Our multi-messenger approach can therefore be extended to such scenarios, offering a broader framework for probing dense DM environments.


%

\clearpage

\setcounter{equation}{0}
\setcounter{figure}{0}
\setcounter{table}{0}

\renewcommand{\theequation}{S\arabic{equation}}
\renewcommand{\thefigure}{S\arabic{figure}}
\renewcommand{\thetable}{S\arabic{table}}

\setcounter{page}{1}
\appendix
\onecolumngrid
\date{\today}

\setcounter{equation}{0}
\setcounter{figure}{0}
\setcounter{table}{0}

\renewcommand{\theequation}{S\arabic{equation}}
\renewcommand{\thefigure}{S\arabic{figure}}
\renewcommand{\thetable}{S\arabic{table}}

\setcounter{page}{1}
\onecolumngrid

\section{Supplementary material}\label{SM}

\subsection{The effect of dark matter on the binary}\label{SM}

{
 Prior to entering the GW dominated regime, the secondary black hole is expected to spiral toward the center due to dynamical friction and hardening interactions with stars and dark matter particles. These mechanisms are crucial in driving the binary to sufficiently small separations, where GW emission takes over and governs the final stages of the inspiral.
Our simulations implicitly assume that, at the initial separation, the binary has already entered this GW-dominated regime, such that other forms of interaction—particularly dynamical friction—are no longer dynamically relevant. 
Such condition would be expected to break down if at the initial BBH separation the dynamical friction timescale is shorter than the gravitational-wave coalescence timescale of 10 years. We now demonstrate that, for the initial binary orbit considered in this work, this condition is in practice never met.


The dynamical friction timescale for a compact object of mass \( m \) orbiting a supermassive black hole of mass \( M \) at radius \( r \) in a dark matter distribution of density \( \rho \) is given by
$
t_{\mathrm{DF}} \sim \frac{v^3}{4\pi G^2 m \rho \ln\Lambda},
$
where \( v = \sqrt{GM/r} \) is the orbital velocity and  the Coulomb logarithm is set to
$\ln \Lambda=10$. Using the representative parameters: \( M = 4 \times 10^6\,M_\odot \), \( m = 120\,M_\odot \), and \( r = 10^{-6}\,\mathrm{pc} \), we find that to reduce $t_{\mathrm{DF}}$ below 10 years, the required density is \( \rho \gtrsim  10^{20}\,M_\odot/\mathrm{pc}^3 \). These densities are  orders of magnitude above the central DM densities expected under plausible astrophysical scenarios
\cite[e.g., see Fig. 1 in Ref.][]{2008PhRvD..78h3506V}.
Thus, we conclude that the effect of the DM on the dynamical evolution of the BBH can be safetly neglected. However, we note that this interaction can give rise to a dephasing of the GW signal, which  provides an additional way of probing the existence of DM \cite[e.g.,][]{Eda:2014kra,2018PhRvD..98b3536K,2022PhRvD.105d3009C}.

\subsection{Influence of early inspiral on the dark matter distribution}

 In our models the 
secondary black hole is placed at an initial radius $a_0$. This radius will be reached as the black hole inspirals toward the center of the galaxy owing dynamical friction, strong scaterring off  stars and DM particles, and GW energy loss. $N$-body simulations show that 
at a radius where the 
enclosed mass in the spike becomes comparable to the mass of the secondary black hole, $m$,
the binary will
cause a significant perturbation to the DM distribution and change its density \cite[e.g.,][]{2021ApJ...922...40D}. 
Thus, we expect our models to remain valid as long as $
M(<a_0)\gtrsim m ~ 
$, with the enclosed DM mass given by
\begin{equation}
M(<r) = \frac{4\pi}{3 - \gamma} \, \rho_0 \, r_0^\gamma \, r^{3 - \gamma},
\end{equation}
where $\rho_0$ and $r_0$ are the density normalization and the characteristic radius at which the spike forms. 
Since our models are scale-free, this effectively translates into a validity condition on the scale parameters  of the DM spike density profile.

For our chosen initial radius $a_0$ and a spike slope of $\gamma = 7/3$, the condition $M(<a_0) \gtrsim m$ implies that the dark matter density at 1 mpc must be  $\gtrsim 5 \times 10^{10}\,M_\odot\,\mathrm{pc}^{-3}$ for $q = 3 \times 10^{-6}$, $\gtrsim 4 \times 10^{11}\,M_\odot\,\mathrm{pc}^{-3}$ for $q = 3 \times 10^{-5}$, and $\gtrsim2 \times 10^{12}\,M_\odot\,\mathrm{pc}^{-3}$ for $q = 3 \times 10^{-4}$. These values can be compared to those typically adopted in dark matter spike models of the Galactic center. For example, Ref.~\cite{2007PhRvD..76j3532V} uses a  density of $2 \times 10^{11}\,M_\odot\,\mathrm{pc}^{-3}$ at 1mpc. While the required densities for lower mass ratios fall within this range, the highest mass ratio would require a density about an order of magnitude larger. Given the substantial uncertainties in the structure and normalization of dark matter spikes, such high densities cannot be ruled out, but they lie at the more extreme end of current theoretical expectations.
    }

\end{document}